\pdfoutput=1

\documentclass[article,pre,10pt,twocolumn,longbibliography]{revtex4-2}
\usepackage{microtype}
\usepackage{graphicx}
\usepackage[colorlinks,linkcolor=blue,urlcolor=blue,citecolor=blue]{hyperref}

\usepackage{caption}
\usepackage{subcaption}
\usepackage{mathptmx}
\usepackage{times}
\usepackage{epsfig,amsopn}
\usepackage{graphicx}
\usepackage{bm,amsmath,amssymb}
\usepackage{natbib}
\usepackage{braket}
\usepackage{float}
\usepackage{enumerate}
\usepackage{hyperref}
\usepackage{comment}
\allowdisplaybreaks[1]
\usepackage{tabularx}

\def\be{\begin{equation}}
\def\ee{\end{equation}}
\def\bse{\begin{subequations}}
\def\ese{\end{subequations}}
\def\bcs{\begin{cases}}
\def\ecs{\end{cases}}
\def\bea{\begin{eqnarray}}
\def\eea{\end{eqnarray}}

\newcommand{\eref}[1]{Eq.~\eqref{#1}}%
\newcommand{\fref}[1]{Fig.~\ref{#1}}%
\newcommand{\frefs}[1]{Figs.~\ref{#1}}%
\newcommand{\frefss}[1]{~\ref{#1}}%
\newcommand{\sref}[1]{Sec.~\ref{#1}}%
\newcommand{\Aref}[1]{Appendix}%
\newcommand{\aref}[1]{Appendix~\ref{#1}}%

\newcommand{\opunit}{\textrm{1}\kern-0.22em\textrm{l}}

\newcommand{\ie}{\textit{i.e.,}\;}

\def\n{\nonumber}

\def\la{\langle}
\def\ra{\rangle}
\def\l{\left}
\def\r{\right}
\def\s{\sum}
\def\b{\beta}
\def\db{\Delta}

\def\p{\partial}
\def\lam{\lambda}

\begin{document}


\title{Multiple transitions in an infinite range $p$-spin random-crystal field Blume Capel model}

\author{{\normalsize{}Santanu Das$^{1, 2}$}
{\normalsize{}}}
\email{santanudas@niser.ac.in}

\author{{\normalsize{}Sumedha$^{1, 2}$}
{\normalsize{}}}
\email{sumedha@niser.ac.in}

\affiliation{\noindent $^{1}$School of Physical Sciences, National Institute of Science Education and Research, Jatni 752050, India}

\affiliation{\noindent $^{2}$Homi Bhabha National Institute, Training School Complex, Anushakti Nagar 400094, India}

\begin{abstract}
We study a $p$-spin model with ferromagnetic coupling and quenched random-crystal fields for $p \ge 3$ for spin-1 systems. We find that the model has lines of first order transitions at finite temperature $(T)$ for all $p \ge 3$. For bimodal distribution of the random-crystal field these lines meet at a \emph{triple point} for weak strength of the crystal field $(\db)$. Beyond a critical strength of $\db$, they do not meet and one of the lines ends at a \emph{critical point} $(T_c)$. Interestingly, we find that on increasing $T$ from $T_c$ keeping other parameters fixed, the system undergoes one more transition which is first order in its character. The system thus exhibits a Gardner like transition for a range of parameters for all finite $p \ge 3$. For $p \to \infty$ the model behaves differently and there is only one random first order transition at $T = 0$.
\end{abstract}

\date{\today}
\maketitle


\section{Introduction}

The disordered $p$-spin models have been studied widely due to their connection with the structural glasses \cite{kirkpatrick1987p,kirkpatrick1987dynamics,moore2002p,kirkpatrick1995disordered}. 
In particular in the $p \rightarrow \infty$ limit, the infinite range $p$-spin model with Ising spins and random couplings, known as the Random Energy Model (REM) \cite{de2006random,derrida1981random,derrida1980random} is exactly solvable and presents a useful setting to test the other methods.

In this paper, we introduce and solve an infinite-range $p$-spin interaction model with ferromagnetic coupling and quenched random-crystal fields. Each spin is an integer spin-$1$ which can take three values ($0, \pm 1$). For $p = 2$ the model is the well known Blume Capel model with the random-crystal fields \cite{santos2015mean,santos2018random,jana2016absence,mukherjee2020emergence}. We call this generalisation the $p$-spin random-crystal field Blume Capel model (pRCBCM). In the absence of the crystal field, for $p = 3$ the model was solved for Ising spins on a triangular lattice by Baxter and Wu and is known as the Baxter-Wu (BW) model \cite{baxter1973exact,baxter1974ising1,baxter1974ising2}. The BW model belongs to the $4$-state Potts universality class \cite{domany1978phase}.
The spin-1 generalization of BW model known as the dilute BW model was first introduced and studied by Kinzel \emph{et. al.} \cite{kinzel1981finite}. The spin-1 BW model with pure crystal field has attracted a lot of recent attention \cite{costa2004phase, dias2017critical, jorge2021entropic, fytas2022universality}. In this paper we report the behaviour of the pRCBCM for any $p \ge 3$, including the $p \to \infty$ limit, for bimodal and Gaussian distributions of the random-crystal field on a fully connected graph.

We calculate the quenched free energy of the pRCBCM using large deviation theory \cite{touchette2009large,den2008large} for arbitrarty distribution of the crystal field. We also calculate the disorder averaged exact ground state. In bimodal distribution (BD) and Gaussian distribution (GD), we find ordered ground state for all strengths of the disorder. For all finite $p$, for BD, there are two ordered phases in the ground state that are separated by a first order transition and for $p \rightarrow \infty$ there is only one ground state. In contrast there is only one ordered ground state for the GD for all $p$.

\begin{figure}[t]
\includegraphics[width=0.7\hsize]{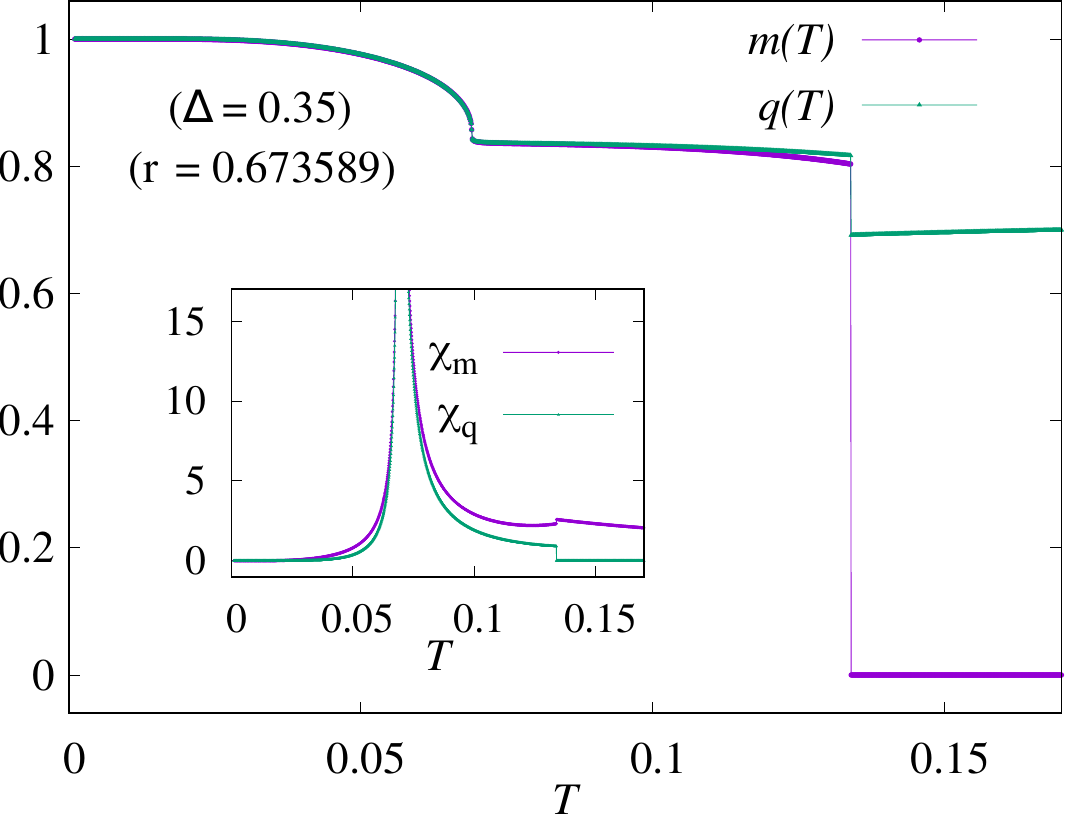}
\caption{(Color online). Magnetization, $m(T)$, and the density of $\pm 1$ spins, $q(T)$,  are plotted versus temperature $T$ for a particular strength of the bimodal distribution $r$ and crystal field strength $\db$ for $p = 3$ to illustrate a continuous transition followed by a first order transition. In the inset susceptibility associated with $m$ and $q$ are plotted versus $T$.}
\label{m_q_gardner_like}
\end{figure}

We find rich phase-diagrams for BD at finite temperatures (T). For $p = 2$ the model has been studied extensively \cite{santos2015mean,santos2018random,jana2016absence,mukherjee2020emergence} and is known to have lines of first order and second order transitions, separated by a tricritical or a critical end-point, depending on the strength of the disorder \cite{mukherjee2020emergence}. For $p \ge 3$ but finite, we find phase diagrams have lines of first order transitions predominantly. Depending on the strength of disorder there are different possible phase diagrams as shown in \frefs{ph_diag_dT} and \frefss{ph_diag_rT}. Interestingly, we also find a Gardner like transition \cite{gardner1985spin} in a narrow range of the parameters, where as one increases $T$ from $0$, the system first undergoes a continuous transition and then a first order transition as shown in \fref{m_q_gardner_like}. This unusual feature in the replica theory corresponds to a transition to a state with full replica symmetry breaking within a 1-RSB state \cite{gardner1985spin}. This kind of transition has been seen in recent experiments with granular glasses \cite{seguin2016experimental}. It has also been reported for $p$-spin glass models \cite{gardner1985spin,gross1985mean} and in the jamming phase diagrams of the granular materials \cite{charbonneau2014fractal,berthier2019gardner}. In the case of pRCBCM, this occurs for all finite $p \ge 3$, though there is no glassy state in the system. Interestingly for $p \to \infty$, there is only one random first order transition (RFOT) that occurs at $T = 0$. In contrast for GD we find only one transition for all strengths of disorder.

The paper is organized as follows. We introduce the pRCBCM in \sref{model}. 
We discuss the phase diagrams for a BD of the random-crystal field in \sref{bimodal} and for GD in \sref{gaussian}. We discuss our result in \sref{summarize}.

\section{Model}
\label{model}

The Hamiltonian of the $p$-spin interacting model in the presence of a quenched random-crystal field is 
\bea
\mathcal{H}(C_N) = - \sum_{1 \le i_1 \le i_2...\le i_p\le N} J_{i_1 i_2...i_p} \;s_{i_1}s_{i_2}...s_{i_p} - \sum_{i=1}^N \db_i s_i^2
\eea 
where $C_N = (s_1, s_2, s_3,...,s_N)$ denotes an arbitrary configuration of $N$ spin variables $s_i$ with the interaction strengths $J_{i_1 i_2...i_p}$ and $\db_i$ is the quenched random-crystal field. For $\db_i = 0$ the above Hamiltonian is a model for spin-glasses \cite{gardner1985spin, mezard1984replica}. Specifically, $p = 2$ is the well-studied Sherrington-Kirkpatrick model of the spin-glass \cite{sherrington1975solvable}.

In this paper we study a model (pRCBCM) with spin-$1$ variables $(s_i = 0, \pm 1)$ and $J_{i_1 i_2...i_p} = 1$. On a fully connected graph, the Hamiltonian of the model becomes  
\bea
\mathcal{H}(C_N)  
= - \frac{1}{p! \;N^{p-1}} \l( \s_{i=1}^N s_i \r)^p - \s_{i=1}^N \Delta_i s_i^2.
\label{H(C_N)}
\eea
For $p = 2$, it is the Hamiltonian of the infinite range random-crystal field Blume Capel model \cite{santos2015mean,jana2016absence,mukherjee2020emergence}.

In this paper, we study a BD of $\db_i$ of the form 
\be
Q(\db_i) =
r \delta(\db_i - \db) + (1 - r) \delta(\db_i + \db),  
\label{rcf_dist}
\ee 
where $r$ and $\db$ are the bias of the distribution and the strength of the random-crystal field respectively. The $r = 0$ and $1$ corresponds to the dilute BW model for $p = 3$ \cite{costa2004phase, dias2017critical, jorge2021entropic, fytas2022universality}. We consider $\db > 0$ and $0  \le r \le 1 $ throughout this paper. Apart from the BD, we also consider a mean-zero GD with the variance $\sigma^2$ for $\db_i$.

This system has two order parameters: the magnetization, $x_1 = \sum_{i} s_i/N$, and the density of $\pm 1$ spins, $x_2 = \sum_i s_i^2/N$ \cite{cardy1996scaling}. For a given sequence of $\{\Delta_i\}$, the probability of a particular configuration $C_N$ can be written as $P_{N, \b} (C_N, \{\db_i\}) = \exp \l[ -\b \mathcal{H}(C_N) \r]/Z_{N, \b}(\{\db_i\})$
with $\b = 1/T$ and the normalization constant, $Z_{N, \beta}(\{\db_i\})$.
It has already been shown in the context of the random-crystal field Blume Capel model that the probability of getting a particular $x_1$ and $x_2$ satisfies large deviation principle (LDP), \cite{jana2016absence} \ie
\be 
\mathbb{P}_{N, \b}\l(C_N: \s_i s_i = x_1 N; \s_i s^2_i = x_2 N\r) \asymp \exp[-N I(x_1, x_2)]
\label{P_N_b_rte_fn}
\ee
where $I(x_1, x_2)$ denotes the rate function, which is like the generalized free energy functional of the model. In the limit of $N \to \infty$, it becomes independent of the specific realization of the disorder on a fully connected graph.

\section{Bimodal random-crystal field}
\label{bimodal}


We calculate the $I(x_1, x_2)$ for pRCBCM defined in \eref{H(C_N)} using the LDP (see \aref{free_energy_a} for details). For a given $\beta$ and $\Delta$, the value of $x_1$ and $x_2$ that minimizes $I(x_1, x_2)$ yields the magnetization $(m)$ and density $(q)$ of the system respectively. Minimizing $I(x_1, x_2)$ with respect to $x_1$ and $x_2$, we obtain a fixed point where $m$ satisfy a self-consistent transcendental equation of the following form
\be
m = \frac{2  r \;e^{\b \db} \sinh \l[ \frac{\beta m^{p-1}}{(p-1)!} \r]}{2 \; e^{\b \db} \cosh\l[ \frac{\beta m^{p-1}}{(p-1)!} \r] +1} + \frac{2 (1- r) \;e^{-\b \db} \sinh \l[ \frac{\beta m^{p-1}}{(p-1)!} \r]}{2 \;e^{-\b \db} \cosh\l[  \frac{\beta m^{p-1}}{(p-1)!} \r] +1}.~~~
\label{m_rcf}
\ee
For a given solution of $m$, $q$ can be expressed in terms of $m$, $\b$, $\db$ and $r$ at the fixed points. The rate function $I(x_1, x_2)$ at the fixed points $(x_1 = m,\; x_2 =q)$ can be expressed as a one parameter functional $\widetilde{f}(m)$, which comes out to be
\bea
\n
\widetilde{f}(m) = \frac{\b m^p}{p(p-2)!} - r\; \log \l( \frac{1 + 2 \; e^{\b \db} \cosh \l[ \frac{\b m^{p-1}}{(p-1)!} \r] }{1+ 2 e^{\b \db}}\r) \\  - (1-r) \;\log \l( \frac{1 + 2 \; e^{-\b \db} \cosh \l[ \frac{\b m^{p-1}}{(p-1)!} \r] }{1+ 2 e^{-\b \db}}\r).~~~~~~~~~
\label{f_m_rcf}
\eea
The value at the minimum of this function and the corresponding $m$ give respectively the free energy and the magnetization of the system. 

\vspace{-0.5cm}
\subsection{Ground state phase diagram}
\label{grund_st_pd}
\vspace{-0.2cm}

At $T = 0$, the disorder averaged ground state energy is given by $\underset{m}{min}\; \phi(m)$, where $\phi(m) = \underset{\b \to \infty}{lim} \b^{-1} \widetilde{f}(m)$. From \eref{f_m_rcf}, for $\Delta >  m^{p-1}/(p-1)!$ it is
\begin{align}
\phi(m) &= \frac{m^p}{p (p-2)!} - \frac{r m^{p-1}}{(p-1)!} 
\end{align}
and for $\Delta < m^{p-1}/(p-1)!$ is
\begin{align}
\phi(m) = \frac{m^p}{p (p-2)!} - \frac{m^{p-1}}{(p-1)!} + (1-r) \db
\end{align}
Taking $\partial \phi(m)/\partial m =0$ we get possible fixed point values as $m=0$ and $m=r$ for $\db > m^{p-1}/(p-1)!$ and $m=0$ and $m=1$ for $\db < m^{p-1}/(p-1)!$. We observe that the ground state is always ordered, with $m=1$ for $\Delta < \Delta_g$ and $m=r$ for $\Delta > \Delta_g$, where $\Delta_g$ is given by
\begin{align}
\Delta_g = \frac{1}{p!} \frac{1-r^p}{1-r}
\label{delta_c}
\end{align}
The model has a first order transition between the two ordered phases given by $m=1$ and $m=r$ at $T=0$ for all finite $p \ge 2$. For $p \rightarrow \infty$, there is only one ordered state with $m=r$ at $T = 0$ as $\db_g\to 0$

\vspace{-0.5cm}
\subsection{Phase diagram for $p \rightarrow \infty$}
\label{p_infty}
\vspace{-0.2cm}

For $p \to \infty$ the model is always in $m = r$ ordered state for all values of $\db$ at $T = 0$. For large $\b$ and large $p$, with $\db > \b m^{p-1}/(p-1)!$ we can take
\begin{align}
\widetilde{f}(m) = \frac{\beta m^p}{p (p-2)!} - r \log \cosh \left(\frac{\beta m^{p-1}}{(p-1)!}\right).
\label{p_inf1}
\end{align}
The magnetisation $m$ is then given by the self consistent equation
\begin{align}
m =r \tanh \left(\frac{\beta m^{p-1}}{(p-1)!}\right).
\label{p_inf2}
\end{align}
If we take the limit $p \rightarrow \infty$ , before taking $\beta$ to infinity, the only fixed point is $m=0$. If we take the $\beta \rightarrow \infty$ first and then $p \to \infty$, then as shown in \sref{grund_st_pd}, the fixed point that minimizes $\widetilde{f}(m)$ is $m=r$. Hence there is a first order transition from $m=0$ to $m=r$ at $T=0$ when $p \rightarrow \infty$. To understand this transition better, we look at the average energy of the model. The average energy (E(m)) which is $\frac{\partial \widetilde{f}(m)}{\partial \beta}$ is given by
\begin{align}
E(m) = \frac{{m}^p}{p (p-2)!}-r \frac{{m}^{p-1}}{(p-1)!} \tanh\left(\frac{\b {m}^{p-1}}{(p-1)!}\right).
\end{align}
For $p \rightarrow \infty$, $E(m)$ is $0$ for all values of $m$. The model has a first order transition at $T=0$ with no latent heat. This puts the transition into the RFOT category \cite{kirkpatrick2015colloquium}. The model has an entropy vanishing transition just like REM \cite{derrida1981random,derrida1980random} but now at $T=0$.

\vspace{-0.5cm}
\subsection{Finite temperature phase diagrams for $p = 3$}
\label{bw_p3}
\vspace{-0.2cm}

For $p = 3$, we determine finite temperature phase diagrams by finding the global minimum of the free energy functional in \eref{f_m_rcf}. We investigate phase diagrams both in $(\Delta-T)$ and $(r-T)$ planes for different $0 \le r \le 1$ and $\Delta > 0$ respectively. 

\vspace{-0.5cm}
\subsubsection{$(\db-T)$ plane}
\label{pd_1}
\vspace{-0.2cm}

As shown in \sref{grund_st_pd}, the ground state for an arbitrary $0 \le r < 1$ has $m \approx 1$ $(r)$ when $\db \le \db_g$ $(\db > \db_g)$, with $\db_g$ given via \eref{delta_c}. 

\begin{figure}[t]
\includegraphics[width=0.85\hsize]{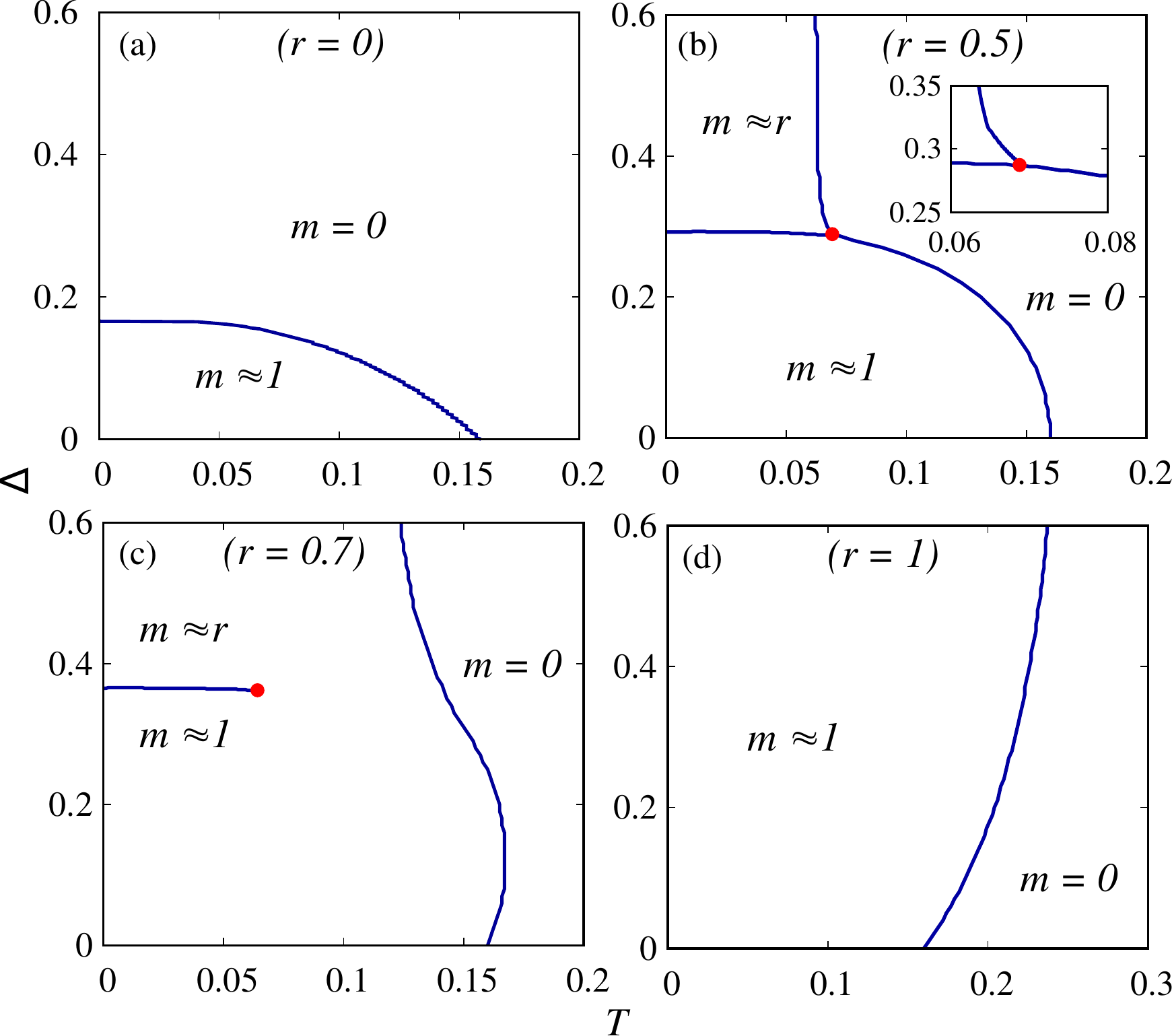}~
\caption{(Color online). Phase diagram at four representative points of $r$ in $(\db-T)$ plane for $p = 3$. Solid lines in each plot represent lines of first order transitions. Solid red circles in (b) and (c)  indicate the triple $(T_t = 0.069000(1)$, $\db_t = 0.286608(2) )$ and critical points $(T_c =0.064269(2)$, $\db_c = 0.361196(6) )$ respectively. In the inset of (b) vicinity of the triple point is highlighted to show the two first order transitions in this regime.} 
\label{ph_diag_dT}
\end{figure}

The ground state behavior gives a cue for the phase-diagram in the entire $(\db-T)$ plane. For $r = 0$ we get a ferromagnetic and a paramagnetic phase separated by a first order transition line. For $0 < r < 1$, we find two  ferromagnetic phases and a paramagnetic phase demarcated by three first order lines of transition.  Interestingly, for $0 < r < 1$, the phase diagrams can be divided in two different categories, $r \le r_*$ and $r > r_*$, with $r_* \approx 0.548(1)$. For $0 < r \le r_*$ three first order transition lines meet at a common point, which is a \emph{triple point} where all three different phases coexist. On the other hand, for $r_* < r < 1$ the first order transition line separating the two ferromagnetic phases terminates at a \emph{critical point} without touching the first order line of transition that demarcates the ferro- and paramagnetic phases. As $r$ approaches $1$, the $T$ associated with this \emph{critical point} approaches zero and vanishes at $r = 1$. The four different phase diagrams are illustrated in \fref{ph_diag_dT} by suitably choosing $r = 0$, $0.5$, $0.7$ and $1$ respectively.

For $0 < r \le 1$, we observe that the first order transition line between ferromagnetic and paramagnetic phases is almost parallel to $\db-$axis in the $(\db-T)$ plane for large $\db$. We evaluate transition temperature $T^*$ by taking the limit of $\db \to \infty$. For a finite $\b$, this limit is equivalent to taking $\db > \b m^{p-1}/(p-1)!$. Hence, Eqs. \eqref{p_inf1} and \eqref{p_inf2} hold also as $\db \to \infty$. These two equations along with the coexistence condition, $\widetilde{f}(m^*) = \widetilde{f}(0) = 0$ gives 
\bea
m^* \l(1 -\frac{1}{p} \r) \tanh^{-1}\l( \frac{m^*}{r}\r) + \frac{r}{2} \log \l( 1 - \frac{{m^*}^2}{r^2} \r) = 0.
\label{db_asymp}
\eea
For a given $r$ and $p$ this equation along with \eref{p_inf2} gives the value of $m^*$ and the corresponding $T^*$ at the first order transition. For example, in case of $r = 0.5$ and $p = 3$ we get $m^* \approx 0.47403$ and $T^* \approx 0.0619$. This matches with the numerical estimates as shown in \fref{ph_diag_dT}(b).

\vspace{-0.5cm}
\subsubsection{$(r-T)$ plane}
\label{pd_2}
\vspace{-0.2cm}

We now study the $(r-T)$ phase diagrams, keeping $\db$ fixed. At $T = 0$ the system is always in a phase with $m = 1$ for $0 < \db \le 1/6$. On the other hand, there are two ferromagnetic phases, $m = 1$ and $r$ within the range $1/6 < \db \le 1/2$. For $\db > 1/2$, again there is only one phase with $m = r$.

\begin{figure}[t]
\includegraphics[width=0.85\hsize]{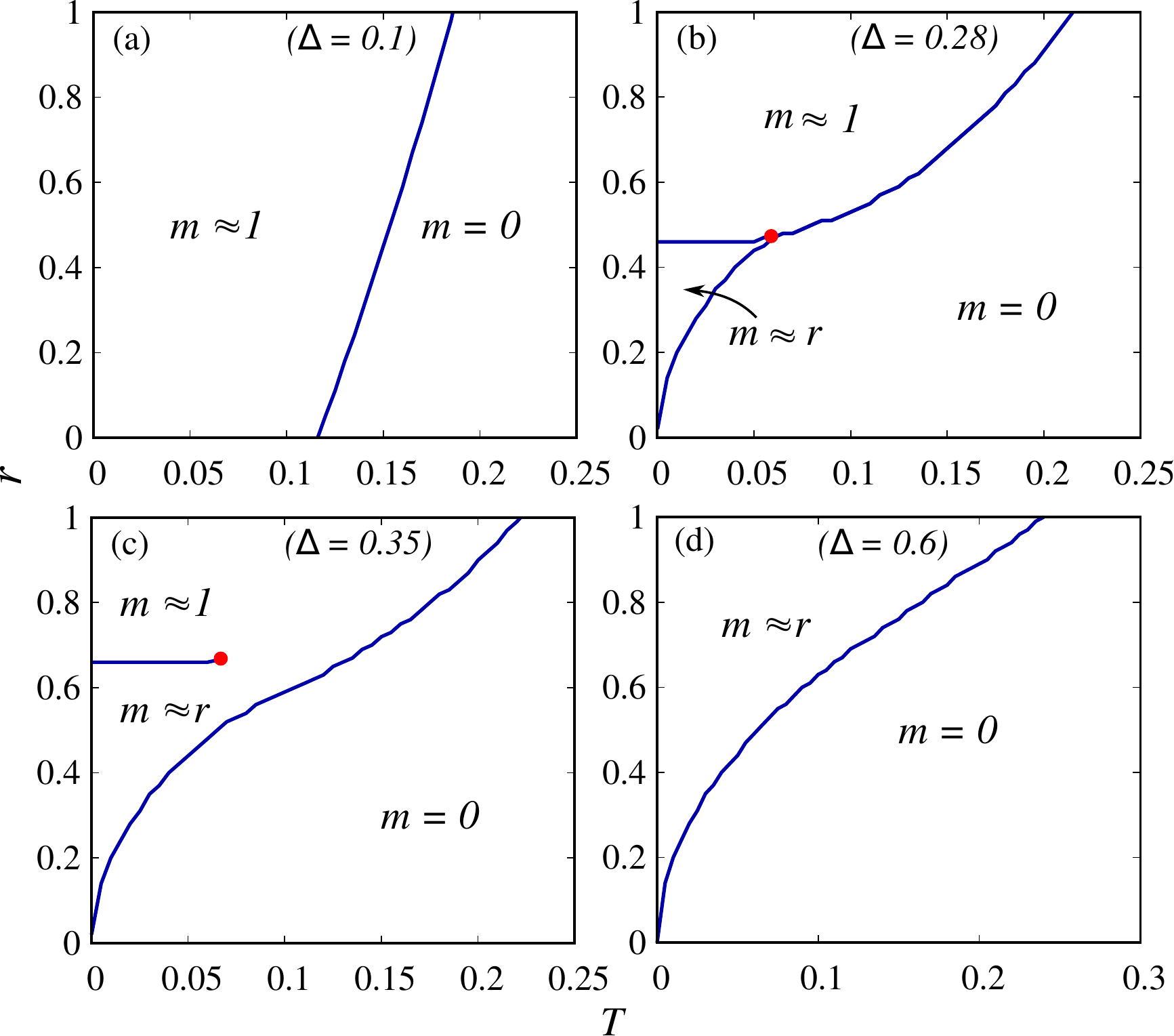}~
\caption{(Color online). Phase diagram at four representative points of $\db$ in $(r-T)$ plane for $p = 3$. Solid lines in each plot represent lines of first order transitions. Solid red circles in (b) and (c) indicate the triple $(T_t = 0.060070(5)$, $r_t = 0.474685(2)) $ and critical $(T_c = 0.069006(9)$, $r_c = 0.673588(9))$ points respectively.}
\label{ph_diag_rT}
\end{figure}

In the range of $0 < \db \le 1/6$, the $(r-T)$ phase diagram consists of a first order transition line separating the two phases, $m \approx 1$ and $m = 0$. For $1/6 < \db \le 1/2$, the qualitative behavior of the phase diagrams in $(r-T)$ plane is the same as that of $0 < r < 1$ in the $(\db-T)$ plane. We observe two different phase diagrams depending on whether $\db \le \db_*$ and $\db > \db_*$, where $\db_* \approx 0.298(1)$. For $1/6 < \db \le \db_*$ we find a \emph{triple point} at the meeting of three first order lines of transition that separate three phases. For $ \db_* < \db \le 1/2$ the first order line that separates the two ferromagnetic phases ends at a \emph{critical point}. The temperature associated with this \emph{critical point} decreases with the increase of $\db$ from $\db_*$, and eventually vanishes for $\db = 1/2$. Above $\db = 1/2$ we hence observe only two phases, $m \approx r$ and $m = 0$. These four different phase diagrams are demonstrated in \fref{ph_diag_rT} by conveniently choosing $\db = 0.1$, $0.28$, $0.35$ and $0.6$ respectively.

\vspace{-0.7cm}
\subsubsection{Multiple transitions}
\label{mult_trans}
\vspace{-0.2cm}

The ferromagnetic to paramagnetic transition for all $p \ge 3$ is always first order. This can be seen by looking at the exact expression of the magnetic susceptibility for $m = 0$ state (see \eref{chi_m0_p3} in the \aref{free_energy_a}).

\begin{figure}[t]
\includegraphics[width=0.7\hsize]{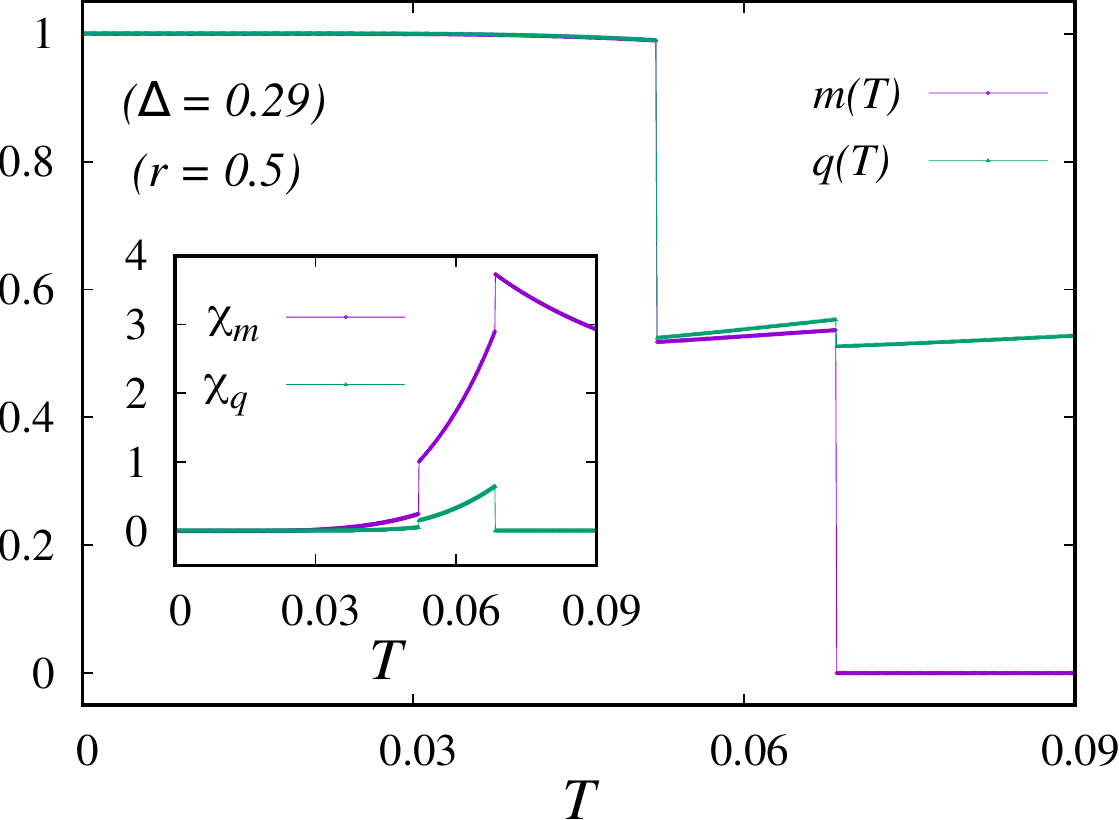}
\caption{(Color online). Magnetization, $m(T)$, and the density $q(T)$,  are plotted versus temperature $T$ for a particular bias of the BD $r$ and crystal field $\db$ for $p = 3$ to illustrate the two first order transitions. In the inset susceptibility associated with $m$ and $q$ are plotted versus $T$.}
\label{m_q_ff}
\end{figure}

In \sref{pd_1} and \sref{pd_2} we noticed that the phase diagram has a \emph{triple point} within the ranges of $0 < r \le r_*$ and $1/6 < \db \le \db_*$. Within these ranges if we choose $\db$ for a given $r $ in such a way that $\db \lesssim \db_g$, then we observe two first order transitions with finite jump in the order parameters and their corresponding susceptibilities as a function of $T$. For example for $r = 0.5$ we get $\db_g = 0.291667$. For $\db = 0.29$ there are two first order transitions as shown in \fref{m_q_ff}.

\begin{figure}[t]
\includegraphics[width=0.8\hsize]{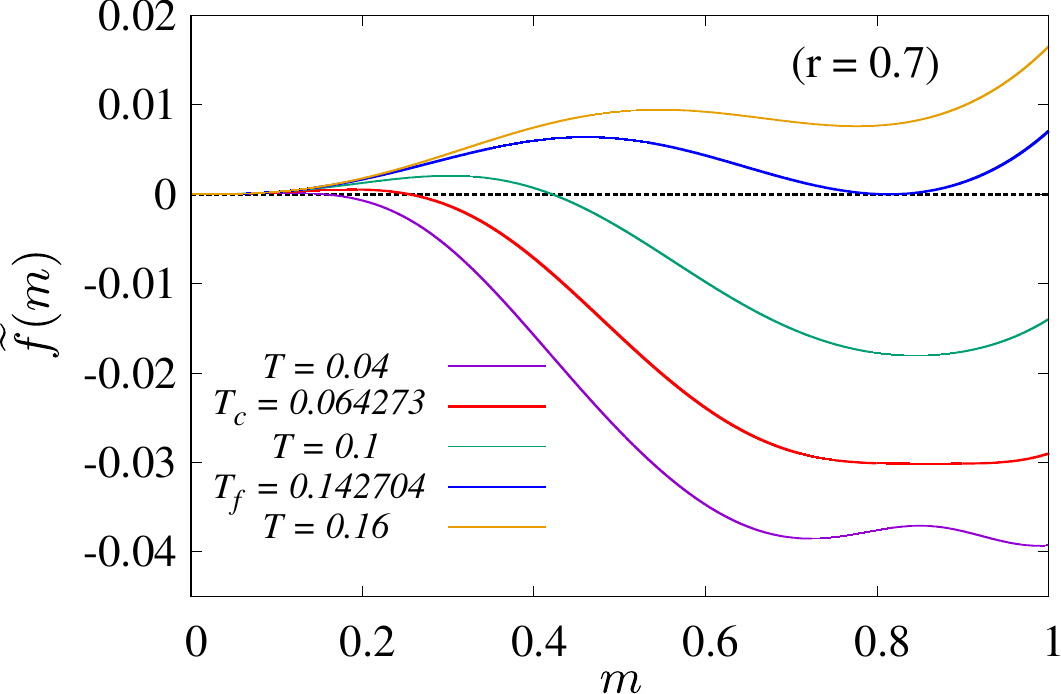}~
\caption{(Color online). Free energy functional $\widetilde{f}(m)$ plotted as a function of magnetization $m$ at different temperatures for $r  = 0.7$ and $\db = 0.361196$ in the case of $p = 3$ for BD.} 
\label{mag_f}
\end{figure}

The phase diagrams for $r_* < r < 1$ and $\db_* < \db \le 1/2$, have a \emph{critical point}. If we go along the critical point by increasing $T$, we observe two different phase transitions. The first one is a continuous transition that occurs at temperature $T_c$ where the order parameter $m$ changes smoothly. In contrast, the second transition is a first order transition at $T_f > T_c$ where $m$ changes abruptly (see \fref{m_q_gardner_like} for details). It is illustrative to look at the free energy functional $\widetilde{f}(m)$ around these transitions. In \fref{mag_f} we plot $\widetilde{f}(m)$ versus $m$ at different $T$s. At the lowest $T = 0.04$ there are two minima with a global minimum of $\widetilde{f}(m)$ at $m \approx 1$. Then at $T_c = 0.064273$ we observe a plateau in $\widetilde{f}(m)$ as shown by the red line. This plateau bears the signature of criticality. At $T_f = 0.142704$, the free energy functional highlighted by the blue line exhibits two minima with $\widetilde{f}(m) = 0$. This $T_f$ is associated with the first order transition as $m \approx r$ and $m = 0$ below and above $T_f$ which can be seen from the plot of $\widetilde{f}(m)$ at $T =0.1$ and $0.16$ respectively. Notably, these successive second and first order transitions of the order parameter $m$ with $T$ are quite similar to the Gardner transition \cite{gardner1985spin} observed in a system of $p$-spin interacting Ising \cite{gardner1985spin}, Potts \cite{gross1985mean} and spin-1 \cite{schelkacheva2015spin} spin glass. The first order transiton in these systems is associated with no latent heat which is sharply in constrast with pRCBCM where we observe a finite latent heat. Apart from that, the second order transition is between the two glassy states in spin glass systems whereas in pRCBCM the transition is between the two ordered states.

\subsection{Finite temperature diagram for other values of $p$}
\label{gen_p}

For $\db = 0$, from \eref{m_rcf} we find that the transition temperature decreases monotonically to $0$ with the increase of $p$. In the limit of $\db \to \infty$ we observe a similar behavior in $T^*$ from Eqs. \eqref{p_inf2} and \eqref{db_asymp}. In fact, in the case of $p = 4$, $5$ we find the qualitative behavior of phase diagrams remain the same as that of $p = 3$. We again observe single or multiple first order transtion lines and a \emph{triple} or a \emph{critical point} within a certain range of parameters. The qualitative behavior of the phase diagram is identical in all cases of $p \ge 3$ for finite $p$, but the area of the ferromagnetic domain within the phase diagrams decreases as we increase $p$. This area goes to $0$ in the limit of $p \to \infty$ where we observe a RFOT between $m = r$ and $0$ at $T = 0$.

For $p = 2$, the model has a different behavior which has been studied earlier \cite{santos2015mean,santos2018random,jana2016absence,mukherjee2020emergence}.  
For $p = 2$ also the \emph{critical point} occurs in the ordered phase, but it is followed by another second order transition at a higher $T$.

\section{Gaussian random-crystal field}
\label{gaussian}

In the case of GD of the form $Q(\db_i) = \exp(- \db_i^2/2 \sigma^2)/\sqrt{2\pi \sigma^2}$, the free energy functional for the model is
\be
\widetilde{f}(m) = \frac{\b m^p}{p(p-2)!} - \int_{-\infty}^{\infty} d\db\; Q(\db) \; \log \l[ \frac{1 + 2 \; e^{\b \db} \cosh \l( \frac{\b m^{p-1}}{(p-1)!} \r) }{1+ 2 e^{\b \db} }\r].
\label{fG_app}
\ee
For $\b \to \infty$ we get
\bea
\n
\phi(m) = \frac{m^p}{p (p-2)!} - \frac{m^{p-1}}{2 (p-1)!} \l( 1 + \text{erf}\l[ \frac{m^{p-1}}{\sqrt{2}\sigma (p-1)!} \r] \r) \\
+ \frac{\sigma}{\sqrt{2 \pi}}\; \l(1 - \exp \l[ - \frac{m^{2(p-1)}}{2 \sigma^2 ((p-1)!)^2} \r] \r). ~~~~~~~~~~
\eea
This gives two equations for $m$ at the fixed point, $m = 0$ and the transcendental equation 
\bea
m = \frac{1}{2} \l( 1 + \text{erf}\l[ \frac{m^{p-1}}{\sqrt{2} \sigma (p-1)!}\r] \r).
\label{trans_m_a}
\eea 
The above equation has a non-zero solution for all values of $p$ and $\sigma$. This non-zero $m$ always minimizes $\phi(m)$. For any finite $p$, \eref{trans_m_a} yields $m = 1$ for $\sigma = 0$. As we increase $\sigma$, the solution smoothly decreases to $m = 1/2$ as $\sigma \to \infty$. Convergence to $m =1/2$ becomes faster with the increase of $p$. In the limit of $p \to \infty$, \eref{trans_m_a} yields $m = 1/2$ for all $\sigma$. Hence, by taking $\b \to \infty$ first, and then $p \to \infty$ we get the ground state magnetization as $m = 1/2$.

For $p \to \infty$ first, and then $\b \to \infty$ we find $m = 0$ from the fixed point equation for $m$, which is obtained by differentiating $\widetilde{f}(m)$ with respect to $m$.
Hence, for $p \to \infty$ we again get a first order transition from $m = 0$ to $m = 1/2$ at $T = 0$. Notably this is identical to the case of BD with $r = 1/2$. Apart from that the average energy $E(m)$ is equal to $0$ for both $m = 0$ and $1/2$. This brings the transition into the RFOT category \cite{kirkpatrick2015colloquium} similar to the case of the BD as discussed in \sref{p_infty}.

\begin{figure}[t]
\includegraphics[width=0.95\hsize]{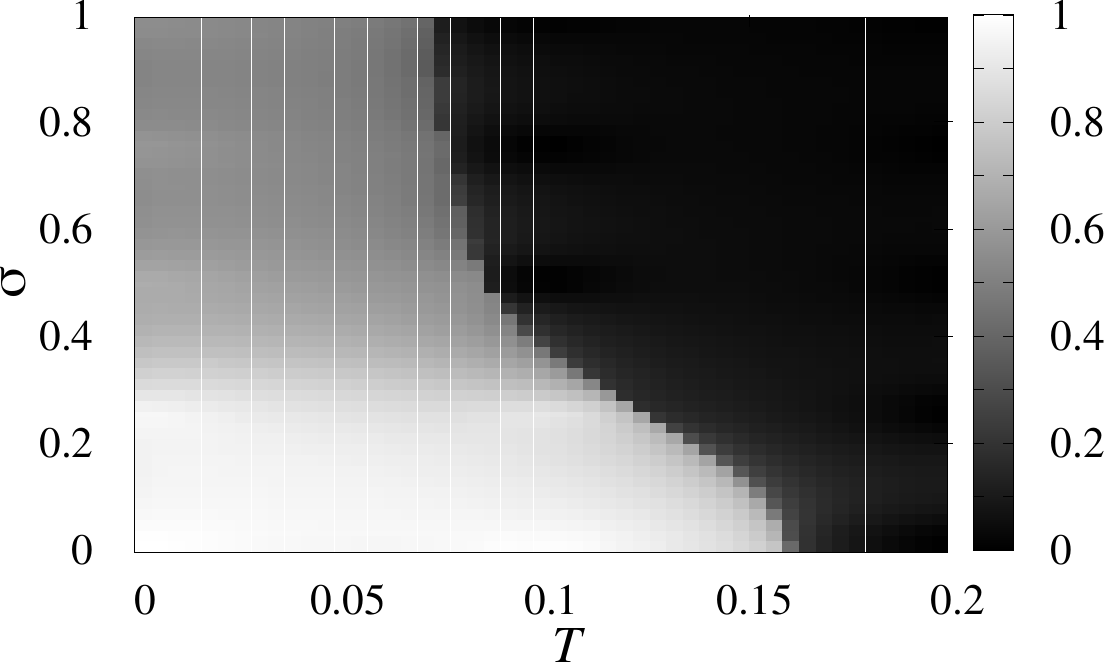}~
\caption{(Color online). Phase diagram of the model in the case of GD plotted  in $(\sigma-T)$ plane for $p = 3$. The density bar on the left side shows the value of $m$.} 
\label{pd_gus}
\end{figure}

For finite $T$ and $p$ we numerically study the global minima of \eref{fG_app} and find that the ferromagnetic and paramagnetic phases are separated by a first order line of transtion in the $(\sigma-T)$ plane. The qualitative behaviour of the phase diagram is identical to that of \fref{ph_diag_dT}(c), albeit without the critical point and the first order transition line separating the two ferromagnetic phases in the later case. The phase diagram of the model calculated numerically by studying $\widetilde{f}(m)$ in \eref{fG_app} is shown in \fref{pd_gus}.

\section{Summary and Discussion}
\label{summarize}

We studied a $p$-spin infinite range ferromagnet with a quenched random-crystal field drawn from a BD, and found that the behaviour of the model is similar for all finite $p \ge 3$. One striking feature of the model is the multiple transitions as a function of $T$. Depending on the value of $r$ and $\db$, either there are two first order transitions or a second order transition followed by a first order transition as $T$ is increased. The high $T$ first order transition in both cases is a result of shrinking of entropy as sites with $s_i^2 = 1$ freeze into $s_i = 1$ state. The second transition in the case of  $\db < \db_*$ reduces entropy to nearly $0$ at a finite $T$, while for $(\db > \db_*)$ there is a gradual change in the configurational entropy. Similar behaviour has been seen in the case of jamming transitions and expected due to the breaking of free energy minimas further into many marginally stable free energy states \cite{charbonneau2014fractal}. In the case of pRCBCM that we have studied, since there is no glassy state, we find the state like this results from the breaking of the free energy minimum into two asymmetric stable minima with $m \ne 0$ through a \emph{critical point}. We expect similar behaviour for any discrete distribution of the random crystal field. pRCBCM due to its simpler free energy landscape is a useful model to explore this unusual low temperature second order transition.

Interestingly, we do not find multiple transitions in the case of the GD. For all $p \ge 3$, there is only one first order transition for any value of $\sigma$. Only for $p \to \infty$ the symmetric BD and the GD have similar phase diagram with a first order transition in the order parameter at $T=0$ with vanishing latent heat.

Recent studies of $p = 2$ quenched random magnetic field ferromagnets \cite{santos2018random,sumedha2022solution,mukherjee2022phase} have revealed a very rich phase-diagram for discrete distributions. We expect a similar study for $p \ge 3$ $p$-spin model would also result in new ordered states and rich phase diagram. A study of the model on a triangular lattice to look for the possibility of multiple transitions in finite dimensions would also be interesting \cite{dias2017critical,jorge2021entropic,fytas2022universality}.

\appendix
\label{appendix}

\section{Calculation of the free energy functional and the magnetic susceptibility}
\label{free_energy_a}

In this section we derive the free energy functional starting from Eq. (2) in the main text with an additional component $\mathcal{H}_{sH} = -  H \sum_{i=1}^N s_i$ in the Hamiltonian. This additional component captures the effect of an external magnetic field $H$ on the system and the Hamiltonian becomes
\bea
\mathcal{H}(C_N)  
= - \frac{1}{p! \;N^{p-1}} \l( \s_{i=1}^N s_i \r)^p - \s_{i=1}^N \Delta_i s_i^2 -  H \sum_{i=1}^N s_i.
\label{Ha(C_N)}
\eea 
To calculate the free energy functional, we first compute the rate function $I(x_1, x_2)$ defined in \eref{P_N_b_rte_fn} in the main text for the order parameters $x_1 = \sum_i s_i/N$ and $x_2 = \sum_i s_i^2/N$. In doing so, we begin with non-interacting part of the Hamiltonian $- \s_{i=1}^N \Delta_i s_i^2 -  H \sum_{i=1}^N s_i$ in \eref{Ha(C_N)}. We first write the scaled cumulant generating function associated with this non-interacting part of the Hamiltonian for a given set of $\{ \db_i \}$ as
\bea 
\lam(k_1, k_2| \{ \db_i \}) = \lim_{N \to \infty} \frac{1}{N} \log \l\la e^{N(x_1 k_1 + x_2 k_2) }\r\ra.
\label{lam_k1_k2_a}
\eea
The angular brackets on the right hand side denote an average over spins $s_i$.
Note that the probabilities of a spin $s_i$ to choose values $\pm 1$ or $0$ for a given $\db_i$ and $H$ are given by
\bea
\mathcal{P}_i(s_i = +1) = \frac{e^{\b(\db_i + H)}}{2 \cosh(\b H) \; e^{\b \db_i} + 1},\\
\mathcal{P}_i(s_i = -1) = \frac{e^{\b (\db_i - H)}}{2 \cosh(\b H) \; e^{\b \db_i} + 1},\\ 
~\mathcal{P}_i(s_i = 0)~~ =~~ \frac{1}{2 \cosh(\b H) \;e^{\b \db_i} + 1}. 
\eea
Taking an average over $s_i$ by using the above probabilities we find 
\be
\lam(k_1, k_2| \{ \db_i \}) = \lim_{N \to \infty} \frac{1}{N} \sum_{i=1}^N \log \l[ \frac{2 e^{k_2 + \b \db_i} \cosh(k_1 + \b H) +1}{ 2 e^{ \b \db_i} \cosh(\b H) +1}\r].
\ee
Apply an averaging over $\{ \db_i \}$ we get 
\bea
\n
\lam(k_1, k_2) 
= g(k_1, k_2) - g(0,0),~~~~~~~~~~~~~~~~~~~
\label{lam_k1_k2}
\eea  
where
\bea
\n
g(k_1, k_2) = r \;\log \l[ 2 e^{k_2 +\b \db} \cosh\l( k_1 + \b H \r) + 1 \r] \\
+ (1 - r) \;\log \l[ 2 e^{k_2 -\b \db} \cosh\l( k_1 + \b H \r) + 1 \r].
\label{f_k1_k2_a}
\eea
This result in \eref{f_k1_k2_a} ensures that $\lam(k_1, k_2)$ in \eref{lam_k1_k2} is finite and differentiable for all finite $\db$, $H$ and $k_1, k_2 \in \mathbb{R}$. It allows us to employ the G\"artner-Ellis theorem \cite{touchette2009large} to find out the rate function
\bea
R(x_1, x_2) = {\underset{k_1, k_2 \in \mathbb{R}}{sup}} \{ k_1 x_1 + k_2 x_2 - \lam(k_1, k_2) \}
\label{r_x1_x2_pa}
\eea
related to the non-interacting part of the Hamiltonian. 
Taking derivatives with respect to $k_1$ and $k_2$ of $(k_1 x_1 + k_2 x_2 - \lam(k_1, k_2))$ and equating both of them to zero, we write the extremum value of $k_1$ and $k_2$ as
\bea
k_1^* = - \b H + \tanh^{-1}\l(\frac{x_1}{x_2}\r),~~\\
k_2^* = \log \l[ \frac{z}{2x_2} \sqrt{x_2^2 - x_1^2}\r].~~~~~~
\eea
Here $z$ is a function of $r$, $\b$ and $\db$ which satisfies the following relation 
\bea
\frac{x_2}{z} = \frac{r\; e^{\b \db}}{1 + z \;e^{\b \db}} + \frac{(1 - r) \;e^{-\b \db}}{1 + z \;e^{-\b \db}}.
\label{x2_z_rc}
\eea
These $k_1^*$, $k_2^*$ and $z$ further gives us  
\bea
\n
R(x_1, x_2) = x_1 \tanh^{-1}\l( \frac{x_1}{x_2}\r)
+  x_2 \log\l[ \frac{z}{2 x_2}\sqrt{x_2^2 - x_1^2}\r]~~~~~~~~~~ \\ \n
- \b H x_1 - r \; \log \l[ \frac{1 + z \; e^{\b \db}}{1+ 2 e^{\b \db} \cosh(\b H)} \r]~~~~~~~~~~~~~~~~~~~
\\ - (1-r) \; \log \l[ \frac{1 + z \; e^{-\b \db}}{1+ 2 e^{-\b \db} \cosh(\b H)} \r].~~~~~~~~~~~~~~~~~~
\eea

Using this expression for $R(x_1, x_2)$ we now compute $I(x_1, x_2)$ defined in \eref{P_N_b_rte_fn} in the maintext by using tilted large deviation principle (LDP) \cite{den2008large}. It is noteworthy that the tilted LDP can generate a new LDP from an old LDP by a change of the probability measure. Precisely, it allows us to write the rate function as 
\bea
I(x_1, x_2) = R(x_1, x_2) - \frac{1}{p!} \b x_1^p.
\label{I_x1_x2_a}
\eea
The values of $x_1$ and $x_2$ that minimize $I(x_1, x_2)$ for a given $\b$, $\db$ and $H$ give the value of magnetization $(m)$ and density $(q)$.
Furthermore the minimum of the rate function in the $(x_1 , x_2)$ plane gives the free energy for a given $\b$, $\db$, and $H$.
Minimizing $I(x_1, x_2)$ with respect to $x_1$ and $x_2$ we get the equations for $m$ and $q$ as
\bea
\label{m_q_reln}
\frac{m}{q} =\tanh \l(\frac{\b m^{p-1}}{(p-1)!}  + \b H\r),~~~~~~~~~~~~~~~~ \\
z = \frac{2}{\sqrt{1 - m^2/q^2}} = 2 \;\cosh\l(\frac{\b m^{p-1}}{(p-1)!} + \b H\r).~~~~~
\eea
These two relations lead us to express $m$
\bea
\n
m = 2 \sinh \l( \b H + \frac{\beta m^{p-1}}{(p-1)!} \r) \l[ \frac{r \;e^{\b \db}}{2 \; e^{\b \db} \cosh\l( \b H + \frac{\beta m^{p-1}}{(p-1)!} \r) +1}  \r. \\ \l. + \frac{(1- r) \;e^{-\b \db}}{2 \;e^{-\b \db} \cosh\l( \b H + \frac{\beta m^{p-1}}{(p-1)!} \r) +1} \r],~~~~~~~~~~~~~
\label{mH_app}
\eea
and $q$ in terms of $m$ and $H$ via \eref{x2_z_rc} as 
\bea
\n
q = 2 \cosh \l( \b H + \frac{\beta m^{p-1}}{(p-1)!} \r) \l[ \frac{r \;e^{\b \db}}{2 \; e^{\b \db} \cosh\l( \b H + \frac{\beta m^{p-1}}{(p-1)!} \r) +1}  \r. \\ \l. + \frac{(1- r) \;e^{-\b \db}}{2 \;e^{-\b \db} \cosh\l( \b H + \frac{\beta m^{p-1}}{(p-1)!} \r) +1} \r].~~~~~~~~~~~~~
\label{qH_app}
\eea
For even values of $p$, the rate function is symmetric around $m = 0$ and \eref{mH_app} holds for $-1 \le m \le 1$ for $H = 0$. For $H = 0$ the right hand side of \eref{m_q_reln} is always positive for any  odd $p$. Hence for all odd $p$ \eref{mH_app} holds for $0 \le m \le 1$. Since the energy of negative $m$ state is always higher for odd $p$, it is sufficient to study the rate function for positive $m$ when $H =0$.

For a given solution of $m$, $q$ is completely determined by $m$, $\beta$, $\db$ and $r$. It further helps to write the rate function or the free energy functional in terms of one parameter $m$ as:
\bea
\n
\widetilde{f}_H(m) = \frac{\b m^p}{p(p-2)!} - r\; \log \l[ \frac{1 + 2 \; e^{\b \db} \cosh \l( \b H + \frac{\b m^{p-1}}{(p-1)!} \r) }{1+ 2 e^{\b \db} \cosh(\b H)}\r] \\  - (1-r) \;\log \l[ \frac{1 + 2 \; e^{-\b \db} \cosh \l( \b H + \frac{\b m^{p-1}}{(p-1)!} \r) }{1+ 2 e^{-\b \db} \cosh(\b H)}\r].~~~~~~~~~
\label{fH_app}
\eea

The obtained result in the presence of $H$ in \eref{fH_app} allows us to get magnetic susceptibility $\chi_m$. Magnetic susceptibility measures the response of the system to an infinitesimal external magnetic field. For a system of magnetization $m$ exposed to a magnetic field $H$, magnetic susceptibility is defined by $\chi_m = \l(\p m/\p H \r)|_{H \to 0}$. To find out this quantity we first recall that the global minima of free energy functional $\widetilde{f}_H(m)$ yields the magnetization of the system \ie where $\p \widetilde{f}_H(m)/\p m = 0$. It gives an equation of the form: 
\bea 
\frac{\b m^{p-1}}{(p-2)!} - g_1(m) - g_2(m) H + \mathcal{O}(H^2) = 0,
\label{mg_sus_p1}
\eea
with 
\bea
\n
g_1(m) = \frac{2 \b m^{p-2}}{(p-2)!} \l[ \frac{(1- r) \;e^{-\b \db} \sinh \left(\frac{\beta  m^{p-1}}{(p-1)!}\right)}{1 + 2 \;e^{-\b \db} \cosh \l( \frac{\b m^{p-1}}{(p-1)!} \r)} \r.\\ \l.
+ \frac{r \;e^{\b  \db} \sinh \left(  \frac{\b m^{p-1}}{(p-1)!} \right)}{2 e^{\beta  \Delta } \cosh \left(  \frac{\b m^{p-1}}{(p-1)!} \right)+1} \r],~~~~~~~~~~~
\label{b1_a}
\eea
and
\bea
\n
g_2(m) = \frac{2 \b^2 m^{p-2}}{(p-2)!} \l[ \frac{(1 - r)\; e^{-\b \db} \left( 2 e^{-\b \db} + \cosh \left( \frac{\b m^{p-1}}{(p-1)!} \right)  \right)}{\left(1 + 2 e^{-\b \db} \cosh \left( \frac{\b m^{p-1}}{(p-1)!} \right)\right)^2} \r. \\ \l. + \frac{r \;e^{\beta  \Delta } \left(2 e^{\beta  \Delta }
+\cosh \left( \frac{\b m^{p-1}}{(p-1)!} \right)\right)}{ \left( 1+ 2 e^{\b \db } \cosh \left( \frac{\b m^{p-1}}{(p-1)!} \right) \right)^2} \r],~~~~~~~~~~
\eea
obtained by a series expansion of $\widetilde{f}'_H(m)$ aronud $H \to 0$.
Taking derivative of \eref{mg_sus_p1} with respect to $H$ first, and then using $H \to 0$ limit give us
\bea 
\label{chi_pa}
\chi_m = g_2(m) \l( \frac{\b (p-1) \;m^{p-2}}{(p-2)!} - \frac{\p}{\p m}g_1(m) \r)^{-1}.
\eea
Interestingly, in the paramagnetic phase \ie where $m = 0$, we find a closed form expression of $\chi_m$ in terms of $\b$ and $\db$ as
\bea
\chi_m = \frac{2 \b}{(p-1)} \l( \frac{r e^{\b \db}}{2 e^{\b \db} + 1} + \frac{(1 - r) e^{-\b \db}}{2 e^{-\b \db} + 1} \r)
\label{chi_m0_p3}
\eea
for all $p \ge 3$. In particular, in case of $p = 2$, it differs from \eref{chi_m0_p3}. In this case it becomes  
\be
\chi_m = \l( \frac{5 + 4 \cosh (\b \db)}{2 \b ((2 r-1) \sinh (\b \db)+\cosh (\b \db) + 2)} - 1 \r)^{-1}.
\label{chi_m0_p2}
\ee
Notably, $\chi_m$ for $p \ge 3$ is always a finite quantity for any $\b$, $r$ and $\db$, whereas it can diverge in case of $p = 2$ for the same parameters. Result in \eref{chi_m0_p3} infers the transition associated with $m = 0$ phase is always a first order transition for all $p \ge 3$ that we discuss in details in Sec. III . On the other hand, in case of $p  = 2$ \eref{chi_m0_p2} implies transition associated with $m = 0$ phase can either be a first order or a continuous phase transition that depends on the other parameters of the model which has already been reported in \cite{mukherjee2020emergence}.



\begin{thebibliography}{99}
\section*{References}

\bibitem{kirkpatrick1987p} T. R. Kirkpatrick and D. Thirumalai,
Physical Review B \textbf{36}, 5388 (1987). 

\bibitem{kirkpatrick1987dynamics} T. R. Kirkpatrick and D. Thirumalai, Physical review letters \textbf{58}, 2091 (1987).

\bibitem{moore2002p} M. Moore and B. Drossel, Physical review letters \textbf{89}, 217202 (2002).

\bibitem{kirkpatrick1995disordered} T. Kirkpatrick and D. Thirumalai, Transport Theory and Statistical Physics \textbf{24}, 927 (1995).

\bibitem{de2006random} L. O. de Oliveira Filho, F. A. da Costa, and C. S. Yokoi, Physical Review E \textbf{74}, 031117 (2006).

\bibitem{derrida1981random} B. Derrida, Physical Review B \textbf{24}, 2613 (1981).


\bibitem{derrida1980random} B. Derrida, Physical Review Letters \textbf{45}, 79 (1980).

\bibitem{santos2015mean} P. V. d. Santos, F. A. da Costa, and J. M. de Araújo, Physics Letters A \textbf{379}, 1397 (2015).


\bibitem{santos2018random} P. Santos, F. da Costa, and J. de Araújo, Journal of Magnetism and Magnetic Materials \textbf{451}, 737 (2018).



\bibitem{jana2016absence} Sumedha and N. K. Jana, Journal of Physics A: Mathematical and Theoretical \textbf{50}, 015003 (2017).

\bibitem{mukherjee2020emergence} Sumedha and S. Mukherjee, Physical Review E \textbf{101}, 042125 (2020).


\bibitem{baxter1973exact} R. Baxter and F. Wu, Physical Review Letters \textbf{31}, 1294 (1973).


\bibitem{baxter1974ising1} R. Baxter, Australian Journal of Physics \textbf{27}, 369 (1974).

\bibitem{baxter1974ising2} R. J. Baxter and F. Wu, Australian Journal of Physics \textbf{27}, 357 (1974).


\bibitem{kinzel1981finite} W. Kinzel, E. Domany, and A. Aharony, Journal of Physics A: Mathematical and General \textbf{14}, L417 (1981).


\bibitem{costa2004phase} M. Costa, J. Xavier, and J. Plascak, Physical Review B \textbf{69}, 104103 (2004).

\bibitem{dias2017critical} D. Dias, J. Xavier, and J. Plascak, Physical Review E \textbf{95}, 012103 (2017).


\bibitem{jorge2021entropic} L. Jorge, P. Martins, C. J. DaSilva, L. Ferreira, and A. Caparica, Physica A: Statistical Mechanics and its Applications \textbf{576}, 126071 (2021).


\bibitem{fytas2022universality} Vasilopoulos A, Fytas NG, Vatansever E, Malakis A, Weigel M., arXiv preprint arXiv:2205.01494. (2022).


\bibitem{domany1978phase} E. Domany and E. K. Riedel, Journal of Applied Physics \textbf{49}, 1315 (1978).


\bibitem{touchette2009large} H. Touchette, Physics Reports \textbf{478}, 1 (2009).


\bibitem{den2008large} F. Den Hollander, \textit{Large deviations}, Vol. 14 (American Mathematical Soc., 2008).


\bibitem{gardner1985spin} E. Gardner, Nuclear Physics B \textbf{257}, 747 (1985).


\bibitem{seguin2016experimental} A. Seguin and O. Dauchot, Physical review letters \textbf{117}, 228001 (2016).


\bibitem{gross1985mean} D. J. Gross, I. Kanter, and H. Sompolinsky, Physical review letters \textbf{55}, 304 (1985).


\bibitem{charbonneau2014fractal} P. Charbonneau, J. Kurchan, G. Parisi, P. Urbani, and F. Zamponi, Nature communications \textbf{5}, 1 (2014).


\bibitem{berthier2019gardner} L. Berthier, G. Biroli, P. Charbonneau, E. I. Corwin, S. Franz, and F. Zamponi, The Journal of chemical physics \textbf{151}, 010901 (2019).


\bibitem{mezard1984replica} D. J. Gross and M. Mézard, Nuclear Physics B \textbf{240}, 431 (1984).


\bibitem{sherrington1975solvable} D. Sherrington and S. Kirkpatrick, Physical review letters \textbf{35}, 1792 (1975).


\bibitem{cardy1996scaling} J. Cardy, \textit{Scaling and renormalization in statistical physics}, Vol. 5 (Cambridge university press, 1996).




\bibitem{kirkpatrick2015colloquium} T. Kirkpatrick and D. Thirumalai, Reviews
of Modern Physics \textbf{87}, 183 (2015).


\bibitem{schelkacheva2015spin} T. Schelkacheva and E. Tareyeva, arXiv:1512.05508 (2015).


\bibitem{sumedha2022solution} Sumedha and M. Barma, Journal of Physics A: Mathematical and Theoretical \textbf{55}, 095001 (2022).


\bibitem{mukherjee2022phase} S. Mukherjee and Sumedha, arXiv:2203.05330 (2022).


\end{thebibliography}
\end{document}